\documentclass[conference]{IEEEtran}
\IEEEoverridecommandlockouts
\usepackage{amsmath,amssymb,amsfonts}
\usepackage{algorithmic}
\usepackage{graphicx}
\usepackage{textcomp}
\usepackage{xcolor}
\def\BibTeX{{\rm B\kern-.05em{\sc i\kern-.025em b}\kern-.08em
    T\kern-.1667em\lower.7ex\hbox{E}\kern-.125emX}}

\newcommand{\norm}[1]{\left\lVert#1\right\rVert}

\graphicspath{{figures/}}

\begin{document}
\title{\LARGE Study of Cloud-Aided Multi-Way Multiple-Antenna Relaying with Best-User Link Selection and Joint ML Detection
}

\author{F. L. Duarte
            and R. C. de Lamare,~\IEEEmembership{Senior Member,~IEEE}
\thanks{Copyright (c) 2015 IEEE. Personal use of this material is permitted. However, permission to use this material for any other purposes must be obtained from the IEEE by sending a request to pubs-permissions@ieee.org.}
\thanks{F. L. Duarte is with the Centre for Telecommunications Studies (CETUC), Pontifical Catholic University of Rio de Janeiro, Brazil, and the Military Institute of Engineering, IME, Rio de Janeiro, RJ, Brazil. e-mail: flaviold@cetuc.puc-rio.br}
\thanks{R. C. de Lamare is with the Centre for Telecommunications Studies (CETUC), Pontifical Catholic University of Rio de Janeiro, Brazil, and the Department of Eletronic Engineering, University of York, United Kingdon. e-mail: delamare@cetuc.puc-rio.br}}

\maketitle
\begin{abstract}
In this work, we present a cloud-aided uplink framework for multi-way multiple-antenna relay systems which facilitates joint linear Maximum Likelihood (ML) symbol detection in the cloud and where users are selected to simultaneously transmit to each other aided by relays. We also investigate relay selection techniques for the proposed cloud-aided uplink framework that uses cloud-based  buffers and physical-layer network coding. In particular, we develop a novel multi-way relay selection protocol based on the selection of the best link,  denoted as Multi-Way Cloud-Aided Best-User-Link (MWC-Best-User-Link). We then devise the maximum minimum distance relay selection criterion along with the algorithm that is incorporated
into the proposed MWC-Best-User-Link protocol.  Simulations show that MWC-Best-User-Link outperforms previous works in terms of  average delay, sum-rate and  bit error rate.
\end{abstract}


\section{INTRODUCTION}
In wireless networks,  the use of cooperative
diversity \cite{f1} can mitigate the signal fading caused by multipath propagation. The Multi-Way Relay Channel (mRC) \cite{f80} includes both a full data exchange model, in which each user receives data from all other users, and the pairwise data exchange model, which is composed by multiple two-way relay channels. The incorporation of mRC with multiple
relays in a system can significantly improve its performance. Considering 5G requirements \cite{f55},  high spectrum efficiency relaying strategies are key due to their excellent performance. The use of a cloud as a central node can leverage the performance of relay techniques as
network operations and services have recently adopted cloud-enabled solutions in communication networks \cite{f100}.
 The  ability to cost-effectively manage  interference  is  one  of  the  main  advantages  of  adopting the cloud network framework \cite{f100}.  In  the Cloud Radio Access Network (C-RAN) architecture, the baseband processing, usually performed locally at each base-station (BS), is aggregated and performed centrally at a cloud processor. This is enabled by high-speed connections, denoted as fronthaul links, between the BSs and the cloud \cite{f100}. This centralized signal processing enables the interference mitigation across all the users in the uplink and downlink. The BSs in the C-RAN are also referred to remote radio heads (RRHs) as their functionality is often limited to transmission and reception of radio signals \cite{f100}.

The mRC has  multiple clusters of users in which each user aims to multicast a single message to all the
other users in the same cluster \cite{f80}. Processing $\mathcal{L}$ users in a cluster corresponds to an $\mathcal{L}$-way information exchange among the
users in the same cluster. A group of $N$ relays facilitates this exchange, by helping all the users in the system. In particular, the mRC pairwise data exchange model ($\mathcal{L}=2$) is formed by multiple two-way relay channels. In Two-Way Multiple-Access Broadcast Channel (MABC) schemes, based on the decode-and-forward (DF) protocol \cite{f40}, the transmission is organized in  two successive phases: 1)  MA phase -  a relay is selected for receiving  and decoding the messages simultaneously transmitted from two users (sources $S_1$ and $S_2$) and physical-layer network coding (PLNC) is performed on the decoded messages; 2)  BC phase - the same  selected relay broadcasts the decoded messages to the two sources. The Two-Way Max-Min (TW-Max-Min) relay selection protocol \cite{f40} has a high performance, when all the channels are reciprocal and fixed during two consecutive time slots (MA and BC phases).  Otherwise, with non reciprocal channels, the performance of relaying strategies can be enhanced by adopting buffer-aided  protocols, in which the relays are able to accumulate data in their buffers \cite{f9, f14}, before sending data to the destination, as in the Multi-Way Max-Link (MW-Max-Link) \cite{f78} protocol for cooperative multi-input multi-output (MIMO) systems, which selects the best links among $K$ pairs of sources (diversity gain equals $2NK$),  using the extended Maximum Minimum Distance (MMD) relay selection criterion \cite{f41, f411}. Furthermore, in \cite{f87}, the Two-Way Max-Link (TW-Max-Link) protocol (a special case of  MW-Max-Link, for a single two-way relay channel ($K=1$)), also using the extended MMD criterion, was presented.  However, cloud-aided  multi-way protocols using the maximum minimum distance relay  selection criterion, for multiple-antenna systems,  in which each cluster has a particular buffer, have not been previously investigated.

In this work, we develop a cloud-aided framework and a Multi-Way Best-User-Link  (MWC-Best-User-Link) protocol for cooperative MIMO systems, with non reciprocal channels, which selects the best links among $K$ pairs of sources (clusters) and $N$ relay nodes. In order to perform signal detection at the cloud and the nodes, we present maximum likelihood (ML) detectors. We then consider the maximum minimum distance criterion and devise a relay selection algorithm for MWC-Best-User-Link.  Simulations illustrate the excellent performance of the proposed framework, the proposed MWC-Best-User-Link protocol and the relay selection algorithm as compared to previously reported approaches.

This paper is structured as follows. Section II describes the system
model and the main assumptions. Section III presents the
proposed MWC-Best-User-Link protocol, the relay selection criterion and algorithm, and analyzes  MWC-Best-User-Link, in terms of pairwise error probability (PEP) and sum-rate.
Section IV illustrates and discusses the simulation results whereas Section V gives the
concluding remarks.

\section{System Description}
\begin{figure}[!h]
\centering
\includegraphics[scale=0.55]{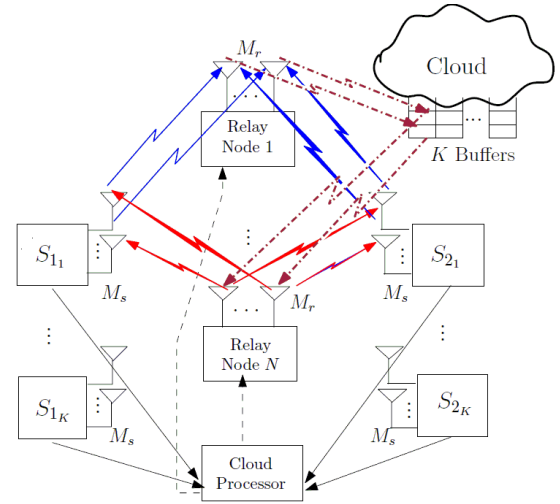}
\caption{System model of the proposed cloud-aided multi-way relay scheme.}
\label{fig:model}
\end{figure}

We assume a MIMO multi-way MABC relay network formed by $K$ clusters (pair of sources $S_1$ and $S_2$) and $N$ half duplex (HD)  DF relays, $R_1$,...,$R_N$.  In a C-RAN framework, the sources would represent mobile users and the relays would represent RRHs. The sources have $M_s$ antennas for transmission or reception and each relay $M_r=2U M_s$ antennas, where $U\in \{1,2,3\dots\}$, all of them used for reception ($M_{r_{Rx}}=M_r$)  and a part of them used for transmission ($M_{r_{Tx}}=VM_s \leq M_r$), where $V\in \{1,2,3\dots\}$, forming a spatial multiplexing network, in which the channel matrices are square or formed by multiple square sub-matrices in the MA  mode. Note that the reason for using multiples of $2M_s$ antennas at the relays is because the relay selection algorithms explained in Section III  use criteria that depend on these matrices to be square or to be formed by multiple square sub-matrices. Moreover, the computational complexity The selected relays access a  number of $K$ cloud  buffers for extracting or storing $M_s$ packets in each time slot. Each cluster has a particular cloud buffer that is established on demand, whose size is $J$ packets, as depicted in Fig.\ref{fig:model}.
In the multiple-access phase (uplink), a cluster is selected to send  $M_s$ packets simultaneously to a selected relay $R_g$ for reception. Then, the data are decoded by the cloud processor, PLNC is employed on the decoded information and the resulting data are stored in their particular cloud buffers. In the broadcast-channel phase (downlink), two relays $R_{j1}$ and $R_{j2}$ are selected to broadcast $M_s$ packets from the particular cloud buffer to the selected cluster. Note that $R_{j2}$ may be different from $R_{j1}$. In most of the situations the selection of only one relay in downlink  is enough for a good performance \cite{f78,f41,f411,f87,f35}. However, by selecting two relays, the possibility of combining the channels related to the selected relays increases the degrees of freedom of the system and, consequently, its performance is improved. The system could select more than two relays to further improve its performance, but the computational complexity would be considerably increased for a high number of relays. For the sake of simplicity, we adopt the mRC pairwise data exchange model, but the full data exchange model can be considered in future works.
\subsection{Assumptions}

The energy transmitted from each source node to the selected relay for reception ($E_s$) or
from the selected relay(s) for transmission to the sources ($E_{r_j}$), in each time slot, is the same, i. e., $E_{r_j}=E_s$. We consider mutually independent
zero mean complex Gaussian random channel coefficients, which are fixed for the duration of one time slot
and vary independently from one time slot to the following, and the transmission is organized in data
packets.  The order of the packets is included in
the preamble and the original order is recovered at
the destination. Signaling for
network coordination and pilot symbols for estimation of
the channel state information (CSI) are also contained in the preamble. The cloud is the central node and decides whether a cluster or the relay(s) must transmit in a given time slot $i$, through a
feedback channel. An appropriate signalling provides global CSI at the cloud. Moreover, we assume that each relay only has information about its $S_1R$ and $S_2R$ links. The use of a cloud as a single central node and its buffers reduces the system complexity and the delay, since a unique central node decides which nodes transmit (rather than all destination nodes) and the packets associated with a cluster are stored in only its particular cloud buffer instead of being spread in the buffers of all relays.
 In this work, we focus on the ideal case where the fronthaul links have unconstrained capacities, and the relays can convey their exact received signals to the cloud processor.  Practical systems, however, have capacity-constrained fronthaul links \cite{f100} and this limits the amount of information that the relays can retransmit. Although these unconstrained capacities in the fronthaul links simplify our analysis, it does not limit the advantages of the proposed protocol and relay selection algorithm, explained in the next section.  Moreover, capacity-constrained fronthaul links can be considered elsewhere in future works and the performance achieved by the proposed protocol may be considered as a baseline or an upper bound.

\subsection{System Model}
For multi-way HD DF MABC  systems, in the MA phase, the signal sent by the selected cluster $S$ ($S_1$ and $S_2$) and received at $R_g$ (the relay selected for reception)  is
organized in an $2UM_s \times 1$ vector given by
\begin{eqnarray}
   \mathbf{y}_{s,r_g}[i]=\sqrt{\frac{E_s}{M_s}} \mathbf{H}_{s,r_g}\mathbf{x}[i]+\mathbf{n}_{r_g}[i],
    \label{eq:2}
\end{eqnarray}
\noindent where $\mathbf{x}[i]$ is
an $2M_s \times 1$ vector with $M_s$ symbols sent
by $S_1$ ($\mathbf{x_1}[i]$) and $S_2$ ($\mathbf{x_2}[i]$),
$\mathbf{H}_{s,r_g}$ is a $2UM_s \times 2 M_s$ matrix of $S_1R_g$ and $S_2R_g$ links and
$\mathbf{ n}_{r_g}$ is the zero mean additive white
complex Gaussian noise (AWGN) at $R_g$. Note that $\mathbf{H}_{s,r_g}$ is formed by $U$ square sub-matrices of dimensions $2M_s \times 2M_s$ as given by
 \begin{eqnarray}
\mathbf{H}_{s,r_g}= [\mathbf{H}^1_{s,r_g}; \mathbf{H}^2_{s,r_g}; \dots ~; \mathbf{H}^U_{s,r_g}].
 \end{eqnarray}

The Maximum Likelihood (ML) detector is the optimal detector from the point of view of minimizing the probability of error (assuming equiprobable $\mathbf{x}$). However, the ML detector has high (exponential in $M_S$) complexity and is only suited to MIMO systems with a small number of antenna elements. Assuming perfect synchronization, we may adopt the ML receiver at the cloud processor:

    \begin{eqnarray}
    \hat{\mathbf{x}}[i]= \arg \min_{\mathbf{x'}[i]} \left(\norm{\mathbf{y}_{s,r_g}[i]- \sqrt{\frac{E_s}{M_s}} \mathbf{H}_{s,r_g}\mathbf{x'}[i]}^2\right),
    \label{eq:4}
    \end{eqnarray}
where $\mathbf{x'}[i]$ is each of the $N_s^{2M_s}$  possible
vectors of sent symbols ($N_s$ is the quantity of  symbols in the constellation adopted). The ML receiver calculates an estimate of the vector of symbols sent by the sources $\hat{\mathbf{x}}[i]$. Other suboptimal detection techniques could be
considered in future work
\cite{mmimo,wence,deLamare2003,itic,deLamare2008,cai2009,jiomimo,Li2011,dfcc,deLamare2013,did,rrmser,bfidd,1bitidd,aaidd}.

By performing PLNC, only the XOR outputs (resulting $M_s$ packets) are stored with the information: "the bit sent by $S_1$ is equal (or not) to the corresponding bit sent by $S_2$". Therefore, we apply the bitwise XOR:
    \begin{eqnarray}
 \mathbf{z}_{[i]}=\mathbf{\hat{x}_1}[i] \oplus \mathbf{\hat{x}_2}[i]
 \end{eqnarray}
 and store the resulting data  in the cloud buffer. Therefore, an advantage of applying PLNC is that we have to store only $M_s$ packets in the cloud buffer, instead of $2M_s$.

In the BC phase, the signal sent by the relays selected for transmission $R_{j}$ ($R_{j_{1}}$ and  $R_{j_{2}}$) and received at $S_1$ and $S_2$ is structured in an $M_s \times 1$ vector
 given by
    \begin{eqnarray}
    \mathbf{y}_{r_j,s_{1(2)}}[i]=\sqrt{\frac{E_{r_j}}{2M_{r_{Tx}}}}  \mathbf{H}_{r_j,s_{1(2)}}\mathbf{z}[i]+\mathbf{n}_{s_{1(2)}}[i],
    \label{eq:3}
    \end{eqnarray}
\noindent where $\mathbf{z}[i]$ is a $M_s \times 1$ vector with $M_s$ symbols, $\mathbf{H}_{r_j,s_{1(2)}}=\mathbf{H}_{r_{j_1},s_{1(2)}}+\mathbf{H}_{r_{j_2},s_{1(2)}}$ represents the $M_s \times M_s$ matrix of  $R_{j_{1}}S_{1(2)}$ and $R_{j_{2}}S_{1(2)}$ links, and $\mathbf{n}_{s_{1(2)}}[i]$ is the AWGN at $S_1$ or $S_2$. Note that $\mathbf{H}_{r_j,s_{1(2)}}$  is formed by summing $V$ matrices of dimension $M_s \times M_s$ as given by
 \begin{eqnarray}
\mathbf{H}_{r_j,s_{1(2)}}= \mathbf{H}^1_{r_j,s_{1(2)}}+\mathbf{H}^2_{r_j,s_{1(2)}}+ \dots + \mathbf{H}^V_{r_j,s_{1(2)}}.
 \end{eqnarray}

We may also adopt the ML receiver at the selected cluster, which yields
    \begin{eqnarray}
    \tilde{\mathbf{z}}_{1(2)}[i]= \arg \min_{\mathbf{z'}[i]} \left(\norm{\mathbf{y}_{r_j,s_{1(2)}}[i]- \sqrt{ \frac{E_{r_j}}{2M_{r_{Tx}}}} \mathbf{H}_{r_j,s_{1(2)}}\mathbf{z'}[i]}^2\right),
    \label{eq:6}
    \end{eqnarray}
where $\mathbf{z'}[i]$  is each of the possible vectors with $M_s$ symbols.

Therefore, at $S_1$ we calculate the vector of symbols sent by $S_2$ by performing PLNC:
 \begin{eqnarray}
\mathbf{\hat{x}_2}[i]= \mathbf{x}_1[i] \oplus \hat{\mathbf{z}}_1[i].
  \end{eqnarray}
It is also applied at $S_2$ to calculate the vector of symbols sent by $S_1$:
 \begin{eqnarray}
 \mathbf{\hat{x}_1}[i]= \mathbf{x}_2[i] \oplus \hat{\mathbf{z}}_2[i].
  \end{eqnarray}
 The estimated channel matrix $\mathbf{\hat{H}}$ is considered instead of $\mathbf{H}$ in (\ref{eq:4}) and (\ref{eq:6}), when performing the ML receiver, by assuming imperfect CSI.  Note that $\mathbf{\hat{H}}$ is computed as $\mathbf{\hat{H}}$=$\mathbf{H}$+$\mathbf{H}_e$, where the variance of the mutually independent zero mean complex Gaussian $\mathbf{H}_e$ coefficients is given by $\sigma_e^2=\beta E^{-\alpha}$ ($0 \leq \alpha \leq 1$ and $\beta \geq 0$) \cite{f16}, in which $E=E_s$, in the MA phase, and $E=\frac{E_{s}}{2}$, in the BC phase. Channel and parameter estimation
\cite{smce,TongW,jpais_iet,armo,badstc,baplnc,goldstein,qian,jio,jidf,jiols,jiomimo,dce}
techniques could be considered in future work in order to develop algorithms for this particular setting.

\section{Proposed MWC-Best-User-Link Protocol and Relay Selection Algorithm}

The system of Fig. \ref{fig:model} is equiped with the novel MWC-Best-User-Link protocol, which in each time slot may operate in two possible modes: MA or  BC. The relay selection algorithm of the proposed MWC-Best-User-Link protocol may operate using the extended MMD \cite{f41} criterion. The MMD-based  relay selection algorithm minimizes the error in the ML receiver  and can be used for  MIMO systems with a small number of antenna elements due to its  reduced complexity in this case.

The MMD-based relay selection algorithm, in the MA mode, chooses the relay and the associated channel matrix $\mathbf{H}_{s,r}^{MMD}$ with the largest minimum distance as given by
\begin{eqnarray}
\mathbf{H}_{s,r}^{MMD}=\arg \max_{\mathbf{H}_{s,r}} \mathcal{B}^{MA}_{min},
  \label{eq:78}
\end{eqnarray}
where $\mathcal{B}^{MA}_{min}=\min \left(\frac{E_s}{M_s}\norm{\mathbf{H}_{s,r}^u(\mathbf{x}_l -
\mathbf{x}_n)}^2\right)$, $u \in \{1, \dots U\}$, $\mathbf{x}_l$  and $\mathbf{x}_n$ represent each possible vector formed by $2M_s$ symbols and $\mathbf{x}_l \neq \mathbf{x}_n$. The metric $\frac{E_s}{M_s}\norm{\mathbf{H}_{s,r}^u(\mathbf{x}_l -
\mathbf{x}_n)}^2$  is calculated for each of the $C_2^{N_s^{2M_s}}$
(combination of $N_s^{2M_s}$  in $2$) possibilities, for each sub-matrix $\mathbf{H}_{s,r}^u$, and $\mathcal{B}^{MA}_{min}$ is the smallest of these values. Thus the selected matrix $\mathbf{H}_{s,r}^{MMD}$ has the largest  $\mathcal{B}^{MA}_{min}$value.
Moreover, the MMD-based relay selection algorithm, in the BC mode, chooses the relay and the associated channel matrix $\mathbf{H}_{r,s}^{MMD}$ with the largest minimum distance as given by
\begin{eqnarray}
\mathbf{H}_{r,s}^{MMD}=\arg \max_{\mathbf{H}_{r,s}} \mathcal{B}^{BC}_{min},
   \label{eq:788}
\end{eqnarray}
where $\mathcal{B}^{BC}_{min} =\min \left(\frac{E_s}{2M_{r_{Tx}}}\norm{\mathbf{H}_{r,s}(\mathbf{x}_l -
\mathbf{x}_n)}^2\right)$, $\mathbf{x}_l$  and $\mathbf{x}_n$
represent each possible vector formed by $M_s$ symbols and $\mathbf{x}_l \neq \mathbf{x}_n$.  The metric $\frac{E_s}{2M_{r_{Tx}}}\norm{\mathbf{H}_{r,s}(\mathbf{x}_l -
\mathbf{x}_n)}^2$  is calculated for each of the $C_2^{N_s^{M_s}}$
 possibilities, for each matrix $\mathbf{H}_{r,s}$, and  $\mathcal{B}^{BC}_{min}$ is the smallest of these values. Thus, the selected matrix $\mathbf{H}_{r,s}^{MMD}$ has the largest  $\mathcal{B}^{BC}_{min}$ value.
The following subsections explain how this protocol works.

\subsection{Relay selection metric for MA and BC modes}

For each cluster $S$ (formed by $S_1$ and $S_2$), in the first step, we calculate the metric $\mathcal{A}^u_{{SR_{i}}}$
related to the $SR$ links of each square sub-matrix $\mathbf{H}^u_{s,r_i}$ associated with the relay $R_i$, in the  MA mode:
\\
\begin{eqnarray}
\mathcal{A}^u_{{SR_{i}}}=  \mathcal{B}_{min}^{MA},
\label{eq:7}
\end{eqnarray}
\\
where $u \in \{1, ...,U\}$ and  $i \in \{1, ...,N\}$.
 In the second step, we compute the ordering on $\mathcal{A}^u_{{SR_{i}}}$  and find the smallest metric, for being critical:
\begin{eqnarray}
  \mathcal{A}_{SR_i} = \min(\mathcal{A}^u_{{SR_{i}}}),
\end{eqnarray}

 In the third step, we compute the ordering on $\mathcal{A}_{SR_i}$ and find the largest metric:
\begin{eqnarray}
  \mathcal{A}_{k_{\max SR}} = \max(\mathcal{A}_{SR_i}),
\end{eqnarray}
where $k \in \{1, ...,K\}$. After finding $\mathcal{A}_{k_{\max SR}}$  for each cluster $k$, we compute the ordering and find the largest metric:
\begin{eqnarray}
  \mathcal{A}_{\max SR} = \max(\mathcal{A}_{k_{\max SR}}).
\label{eq:999}
\end{eqnarray}

Therefore, we choose the cluster and the relay $R_i$ that fulfil (\ref{eq:999}) to receive $M_s$ packets from the selected cluster. For each cluster, in the fourth step, we calculate the metrics $\mathcal{A}_{R_{ij}S_1}$
related to the $RS_1$ links of each matrix $\mathbf{H}_{r_{ij},s_1}$  associated with each pair of relays $R_i$ and $R_j$, for BC mode: \\
\begin{eqnarray}
\mathcal{A}_{R_{ij}S_1}= \mathcal{B}_{min}^{BC},
\label{eq:77}
\end{eqnarray}
\\
where $\mathbf{H}_{r_{ij},s_1}=\mathbf{H}_{r_{i},s_1}+\mathbf{H}_{r_{j},s_1}$, $i$ and $j$ $\in \{1, ...,N\}$. In the fifth step, this reasoning is also applied to calculate the metric $\mathcal{A}_{R_{ij}S_2}$. In the sixth step, we compare the metrics $\mathcal{A}_{ R_{ij}S_1}$ and $\mathcal{A}_{ R_{ij}S_2}$  and store the smallest one:
\begin{eqnarray}
\mathcal{A}_{\min R_{ij}S} = \min(\mathcal{A}_{ R_{ij}S_1},\mathcal{A}_{ R_{ij}S_2}).
\end{eqnarray}

In the seventh step, after finding $\mathcal{A}_{ \min R_{ij}S}$ for each pair of relays, we compute the ordering and find the largest metric:
\begin{eqnarray}
\mathcal{A}_{k_{\max  RS}}=\max(\mathcal{A}_{\min R_{ij}S}),
\end{eqnarray}
where $k \in \{1, ...,K\}$. After finding $\mathcal{A}_{k_{\max  RS}}$ for each cluster $k$, we compute the ordering and find the largest metric:
\begin{eqnarray}
 \mathcal{A}_{\max  RS} = \max(\mathcal{A}_{k_{\max RS}}).
 \label{eq:888}
\end{eqnarray}

 Therefore, we select the cluster and the relays $R_i$ and $R_j$ that fulfil (\ref{eq:888}) to send simultaneously $M_s$ packets stored in the particular cloud buffer to the selected cluster. The estimated channel matrix $\mathbf{\hat{H}}$ is considered in (\ref{eq:7}) and (\ref{eq:77}), instead of $\mathbf{H}$, if we consider imperfect CSI. Alternatively, a designer can consider precoding techniques
\cite{lclattice,switch_int,switch_mc,gbd,wlbd,mbthp,rmbthp,bbprec,baplnc,memd}
to help mitigate interference rather than open loop transmission.

\subsection{Choice of the transmission mode}
After calculating the metrics related to the SR and RS
links and finding $\mathcal{A}_{\max SR}$ and $\mathcal{A}_{\max RS}$, these metrics are compared and we select the transmission mode:\\\\
$
\begin{cases}

             \mbox{if}  ~ \frac{N_{packets}}{M_s} > LoL, ~ \mbox{then}~&\mbox{" BC mode" and select the cluster},\\

               & \mbox{       whose buffer is fullest.}\\

            \mbox{elseif}  ~\frac{\mathcal{A}_{\max SR}}{\mathcal{A}_{\max RS}} \geq G, ~ \mbox{then}  &  \mbox{" MA mode",}\\

             \mbox{otherwise,} & \mbox{" BC mode"},

\end{cases}
$
where $G =\frac{E[\mathcal{A}_{\max  SR}]}{E[\mathcal{A}_{\max  RS}]}$, $N_{packets}$ is the total number of packets stored in the cloud buffers, $LoL\in \{0, 1, 2, \dots\}$ is a parameter that when reduced increases the probability of the protocol to operate in BC mode and, consequently, achieve a reduced average delay (low latency).

\subsection{Pairwise Error Probability}
The PEP assumes an error event when $\mathbf{x}_n$  is sent and the detector calculates an incorrect $\mathbf{x}_l$  (where $l$ $\neq$ $n$), based on the received symbol \cite{f78, f41, f411}. Considering  $\mathcal{D'}=\norm{\mathbf{H}(\mathbf{x}_n-\mathbf{x}_l)}^2$,  in MA mode, and $\mathcal{D'}=\frac{1}{2}\norm{\mathbf{H}(\mathbf{x}_n-\mathbf{x}_l)}^2$, in BC mode, the worst value of the PEP (PEP worst case) that occurs for the smallest value of $\mathcal{D'}$  ($\mathcal{D'}_{\min}$) is given by
\begin{eqnarray}
\begin{split}
\mathbf{P}(\mathbf{x}_n \rightarrow \mathbf{x}_l | \mathbf{H})&= Q\left(\sqrt{\frac{E_s}{2 N_0M} \mathcal{D'}_{\min}} \right),
  \label{eq:19}
\end{split}
\end{eqnarray}
where $M=M_s$, in the MA mode, and $M=M_{r_{Tx}}$, in the BC mode.
By considering that the probability of having no error in the two phases of the system is approximately given by the square of $(1-\mathbf{P}(\mathbf{x}_n \rightarrow \mathbf{x}_l | \mathbf{H}))$, an expression for calculating the worst case of the PEP for cooperative transmissions (CT), in each time slot is given by
\begin{eqnarray}
\begin{split}
\mathbf{P}^{CT}(\mathbf{x}_n \rightarrow \mathbf{x}_l | \mathbf{H})&\approx1-(1-\mathbf{P}(\mathbf{x}_n \rightarrow \mathbf{x}_l | \mathbf{H}))^2\\
&\approx 1- \left(1-Q\left(\sqrt{\frac{E_s}{2 N_0M} \mathcal{D'}_{\min}}\right)\right)^2.
  \label{eq:102}
\end{split}
\end{eqnarray}
Note that this expression may be used for calculating the worst case of the PEP, for both symmetric and asymmetric channels.
The proposed MWC-Best-User-Link, using the MMD relay selection criterion, selects the channel matrix $\mathbf{H}^{MMD}$, minimizing the PEP worst case, as shown by
\begin{eqnarray}
\begin{split}
\mathbf{H}^{MMD}&=\arg \min_{\mathbf{H}} \mathbf{P}(\mathbf{x}_n \rightarrow \mathbf{x}_l | \mathbf{H})\\
&=\arg \max_{\mathbf{H}} \min \norm{\mathbf{H}(\mathbf{x}_n-\mathbf{x}_l)}^2.\\
\end{split}
\end{eqnarray}

Consequently, the MMD relay selection criterion,  by maximizing the minimum Euclidian distance between different vectors of transmitted symbols, minimizes the error in the ML receiver.
This reasoning may be applied also for each of the square sub-matrices $\mathbf{H}^u$ in a non square matrix $\mathbf{H}$ (formed by multiple square sub-matrices). In a future journal version of this paper we develop a  proof that shows that the MMD relay selection criterion minimizes the PEP worst case and, consequently, the error  in the proposed MWC-Best-User-Link protocol, with ML receiver.

\subsection{Sum-Rate}

In \cite{f78}, a framework is proposed to analyze the sum-rate of the MW-Max-Link. In the following, we use this framework to compute the sum-rate of the proposed MWC-Best-User-Link.
In the case of a time slot $i$ selected for MA mode, the sum-rate is given by
\begin{eqnarray}
R_i^{SR}= \frac{1}{2}~  \log_2 \det \left(\mathbf{H}_{s,r} (\mathbf{Q}_{s,r}/N_0)\mathbf{H}_{s,r}^H+\mathbf{I}\right),
  \label{eq:35}
\end{eqnarray}
where $\mathbf{Q}_{s,r}=E[\mathbf{x}(\mathbf{x})^H]= \mathbf{I}~\frac{ E_s}{M_s}$. Furthermore, in the case of a time slot $i$ selected for BC mode, the sum-rate is given by
\begin{eqnarray}
R_i^{RS_{1(2)}}= \frac{1}{2}~  \log_2 \det \left(\mathbf{H}_{r,s_{1(2)}} (\mathbf{Q}_{r,s_{1(2)}}/N_0)\mathbf{H}_{r,s_{1(2)}}^H+\mathbf{I}\right),
  \label{eq:36}
\end{eqnarray}
where $\mathbf{Q}_{r,s}=\mathbf{I}~ \frac{E_s}{2M_{r_{Tx}}}$. So, the average sum-rate ($R$)  of the MWC-Best-User-Link scheme can be approximated by
\begin{eqnarray}
R\approx \frac{\sum_{i=1}^{n_{SR}} R_i^{SR}+\sum_{i=1}^{n_{RS}} (R_i^{RS_1}+ R_i^{RS_2})}{n_{SR}+n_{RS}},
  \label{eq:38}
\end{eqnarray}
where $n_{SR}$ and $n_{RS}$ are the number of time slots selected for SR and RS transmissions, respectively.

\section{Simulation Results}

We assess via simulations the proposed MWC-Best-User-Link and the existing MW-Max-Link \cite{f78}, using the MMD-based relay selection algorithm, with the ML receiver. We employ BPSK signals and note that other constellations as QPSK and 16-QAM were not included but can be examined elsewhere. The average delay is calculated by considering the time a packet needs to reach the destination once it has left the source (no delay is measured when the packet resides at the source \cite{f23}). So, the delay is the number of time slots the packet stays in the cloud buffer. The performance of MWC-Best-User-Link and MW-Max-Link protocols was assessed for a set of $L$ values. Then, we found that $L=\frac{J}{M_s}=3$ sets of $M_s$ packets is sufficient to ensure a good performance.  We consider perfect and imperfect CSI and symmetric unit power channels ($\sigma_{ s,r}^2$ $=$ $\sigma_{ r,s}^2$ $= 1$). The signal-to-noise ratio (SNR) given by $E/N_0$ ranges
from 0 to 10 dB, where $E$ is the energy transmitted from each source or the relay(s) and  we consider $N_0 =1$. The transmission protocols
were simulated for $10000M_s$ packets, each with $T=100$ symbols. We assumed perfect signaling between the cloud and the network, but  imperfect signaling can be considered in future works.

\begin{figure}[!h]
\centering
\includegraphics[scale=0.43]{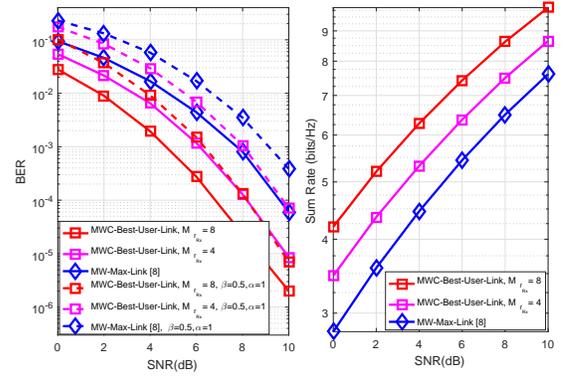}
\caption{BER and Sum-Rate performances versus SNR.}
\label{fig:pepMaxlinkmmse}
\end{figure}

Fig. \ref{fig:pepMaxlinkmmse} depicts the BER and sum-rate performances of the MWC-Best-User-Link (MMD) and MW-Max-Link (MMD)  protocols, for $M_s = 2$, $M_{r_{Tx}}=2$, $M_{r_{Rx}}=4$ in MW-Max-Link and $M_{r_{Rx}}=4$ and $8$ in MWC-Best-User-Link, $K = 5$, $N = 10$, BPSK, $LoL>KL$, perfect and  imperfect CSI  ($\beta=0.5$ and $\alpha=1$).  For both perfect  and imperfect CSI (full and dashed curves, respectively), the BER performance of  MWC-Best-User-Link  is considerably better than that of  MW-Max-Link for all the range of SNR values simulated. Note that the BER performance of MWC-Best-User-Link, with $M_{r_{Rx}}=8$, obtains a gain of almost 3dB in SNR for the same BER as compared to that of MW-Max-Link.  Moreover, the sum-rate performances of
 MWC-Best-User-Link are also considerably better than that of  MW-Max-Link.
\begin{figure}[!h]
\centering
\includegraphics[scale=0.48]{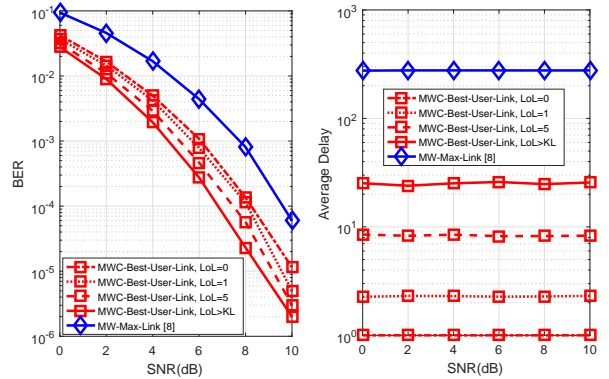}
\vspace{-10pt}
\caption{BER and Average Delay performances versus SNR.}
\label{fig:berAdmmse}
\end{figure}

Fig. \ref{fig:berAdmmse} illustrates the BER and the average delay performances  of MWC-Best-User-Link (MMD) and  MW-Max-Link (MMD), for BPSK, $M_s = 2$, $M_{r_{Tx}}=2$, $M_{r_{Rx}}=4$  in MW-Max-Link, and $M_{r_{Rx}}=8$ in MWC-Best-User-Link, $K=5$, $N = 10$, $LoL=0$, 1, 5 and $LoL>KL$ and perfect CSI. The average delay performance of MWC-Best-User-Link  is considerably better than that of  MW-Max-Link, as  MWC-Best-User-Link has a unique set of $K$ cloud buffers. When we reduce the value of $LoL$ to 0  in the MWC-Best-User-Link protocol, the average delay is reduced to $1$ time slot, still keeping a considerably better BER performance than that of MW-Max-Link.

\section{Conclusions}

A novel framework configured by a cloud as a central node with buffers has been introduced and investigated as a favorable relay selection strategy for multi-way protocols. We have examined relay-selection techniques for multi-way cooperative
MIMO systems that are aided by a cloud central node, where a cluster with two sources is selected to  simultaneously transmit to each other aided by relays. Simulations illustrate the excellent performance of the proposed MWC-Best-User-Link protocol, that by using the MMD-based relay selection algorithm, outperformed the existing MW-Max-Link scheme in terms of BER, sum-rate and average delay. In particular, this novel protocol has a considerably reduced average delay, keeping the high diversity gain.

\end{document}